\documentstyle[epsfig,aps,preprint]{revtex}
\tighten
\newcommand{\n}{\noindent}

\begin{document}

\title{Quantum Chaos of a particle in a square well : Competing Length Scales
and Dynamical Localization}

\author {R. Sankaranarayanan\footnote{sankar@prl.ernet.in},
A. Lakshminarayan\footnote{arul@prl.ernet.in} and
V. B. Sheorey\footnote{sheorey@prl.ernet.in}}
\address{Physical Research Laboratory,\\
Navrangpura, Ahmedabad 380 009, India.}
\maketitle

\begin{abstract}

The classical and quantum dynamics of a particle trapped in a one-dimensional 
infinite square well with a time periodic pulsed field is investigated. This 
is a two-parameter non-KAM generalization of the kicked rotor, which can be 
seen as the standard map of particles subjected to both smooth and hard 
potentials. The virtue of the generalization lies in the introduction of 
an extra parameter $R$ which is the ratio of two length scales, namely the well 
width and the field wavelength. If $R$ is a non-integer the dynamics is 
discontinuous and non-KAM. We have explored the role of $R$ in controlling 
the localization properties of the eigenstates. In particular the connection 
between classical diffusion and localization is found to generalize reasonably 
well. In unbounded chaotic systems such as these, while the nearest neighbour
spacing distribution of the eigenvalues is less sensitive to the nature of 
the classical dynamics, the distribution of participation ratios of the 
eigenstates proves to  be a sensitive measure; in the chaotic regimes the
latter being lognormal. We find that the tails of the well converged 
localized states are exponentially localized despite the discontinuous 
dynamics while the bulk part shows fluctuations that tend to be closer to 
Random Matrix Theory predictions. Time evolving states show considerable $R$ 
dependence and tuning $R$ to enhance classical diffusion can lead to 
significantly larger quantum diffusion for the same field strengths, an 
effect that is potentially observable in present day experiments.

\end{abstract}

\vspace*{1cm}

\newpage
\section {Introduction}

For several years now studies on quantized chaotic systems have
increased significantly with the object of revealing quantum
mechanical manifestations of classical chaos \cite{haake,stockmann}.
The bulk of the work has used smooth Hamiltonian systems.  If we start
perturbing an integrable system, classical Hamiltonian chaos may
develop through a gradual destruction of invariants. The celebrated
Kolmogorov-Arnold-Moser 
(KAM) theorem gives conditions as to when a given tori would be only
distorted. This scenario has been widely studied in two degree of
freedom systems or two-dimensional area preserving maps. However,
there are conditions upon which the KAM theorem rests that may not
always be satisfied by certain systems of physical interest.  In
particular if the perturbation is not sufficiently smooth or even
discontinuous the KAM scenario may break down. Large scale chaos may
instantaneously develop in the system. One other way is that the KAM scenario
fails when the unperturbed system is fully resonant, as in the
Kepler problem. We deal in this paper with the former kind of non-KAM
behaviour.

We first discuss the prevalence of systems where such effects may be
seen. The simplest systems where Hamiltonian chaos can develop is the 
so called $1.5$ degree of freedom system, which are time-dependent 
one-degree of freedom systems. Thus consider the rotor Hamiltonian:

\[ H\,=\, \frac{p_\theta^2}{2}\, +\, f(t)\, V(\theta ) \]

\n where $V(\theta )$ is an external potential that is periodic with 
period $2\pi$, and $f(t)$ is a periodic function of time with period $T$. 
A lengthening pendulum for instance may be the system under study. If 
$V(\theta)$ is sufficiently smooth, the KAM theorem scenario combined with 
the Poincare-Birkhoff theorem provides the generic behaviour. The smoothness 
or at least continuity of $V(\theta)$ is provided by the periodic boundary 
conditions in angular position of the rotor. Introducing discontinuous 
potentials will lead to delta function forces equivalent to walls of certain 
heights. 

This brings us to a natural class of systems where non-KAM behaviour will
be the rule rather than the exception: externally forced particles in wells.
This forms a broad class of systems which have evoked considerable interest
and research since the development of quantum wells and dots. One of the 
experiments where quantum ``scarring'' of wave functions was reported involved
resonant tunneling of a particle across a well in which there were external
electromagnetic fields \cite{wilk96}. 

In fact the simplest of such systems involve a particle in one-dimensional 
infinite square wells (1-d billiards) with time-dependent external
fields. Consider as an example the Hamiltonian:

\begin{equation}
H = H_0 + \epsilon \cos(\omega t) \cos(2 \pi x/ \lambda )
\label{h}
\end{equation}

\n where $H_0 = p^2/2 + V_{sq}(x;a)$, describing such a particle. 
The potential $V_{sq}(x;a)$ is the confining infinite square well potential of 
width $2a$, centered at the origin. Here $\epsilon$ and $\lambda$ are 
field strength and wavelength of the external field which is being modulated
in time with frequency $\omega$. 

It is easy to verify that the equations of motion are invariant under
the following transformation:

\[ t \rightarrow \omega_0 t,\; p\rightarrow p/2a \omega_0,\; 
x\rightarrow x/2a,\]
and 
\[ \epsilon\rightarrow \epsilon/(2a\omega_0)^2, \; 
\lambda\rightarrow\lambda/2a, \; \omega\rightarrow \omega/\omega_0 \,.\]

\n Here the frequency $\omega_0$, which sets the new time scale, is arbitrary. 
Note that the new scaled variables and parameters are referred by the old
symbols and they are dimensionless. Setting $\omega_0 = \omega$ in the
above transformation, we have effectively two parameters : $\epsilon$ and
$R=2a/\lambda$. Here $R$ is the ratio of two length scales of the system
i.e., well width and field wavelength. The presence of two competing length 
scales, provides a rich range of non-KAM behaviours. In particular if the 
dimensionless ratio $R$ is a non-integer there is a possibility of observing 
non-KAM phenomena. 

\begin{figure}
\centerline{\psfig{figure=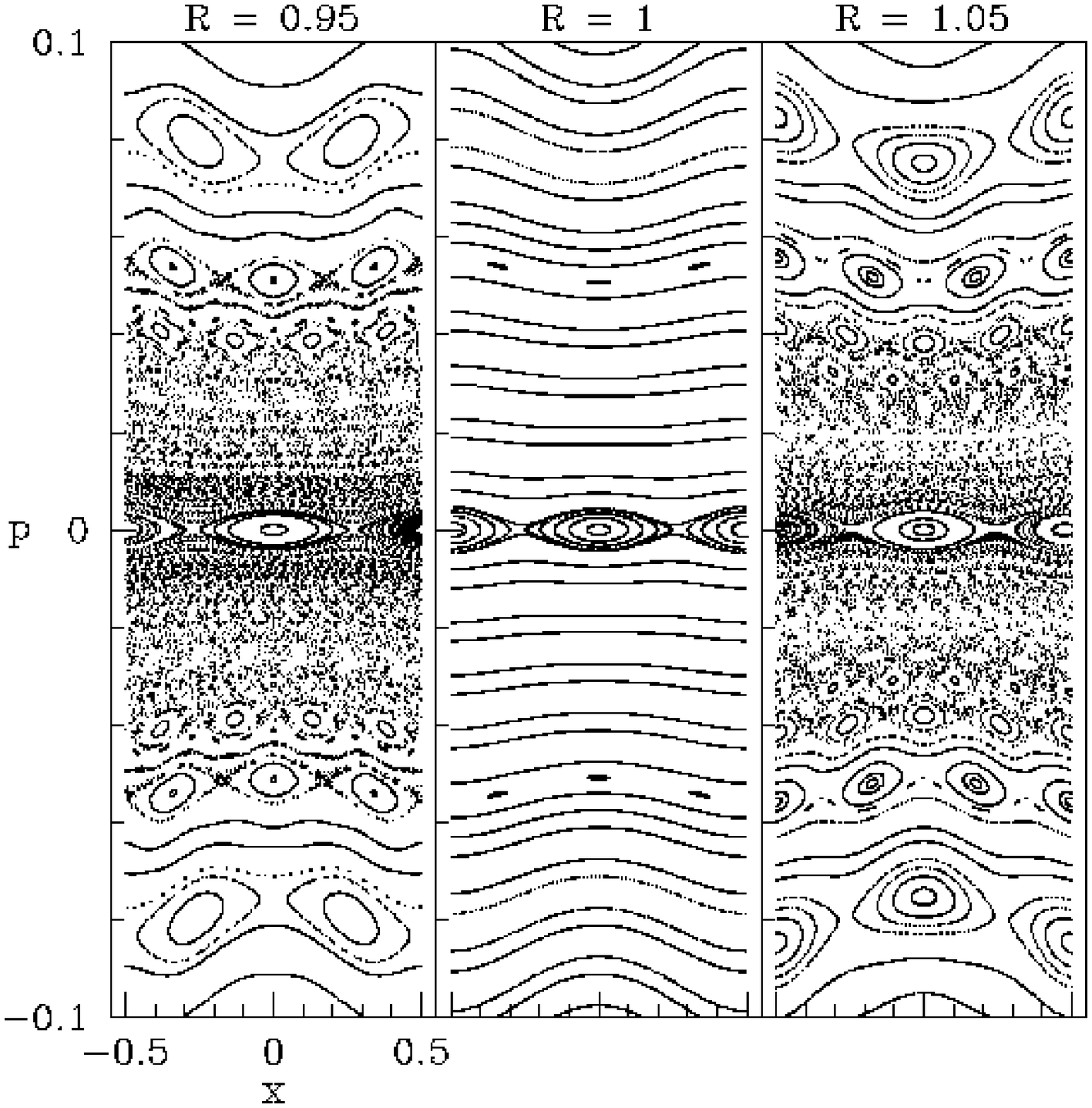,height=15cm,width=15cm}}
\caption{Typical phase space of the system governed by the Hamiltonian in
(\ref{h}) with $\epsilon = 0.001$ and  $\omega=1$. The lower momentum region 
is increasingly chaotic when the length scales do not match.}
\label{xp}
\end{figure}

\begin{figure}
\centerline{\psfig{figure=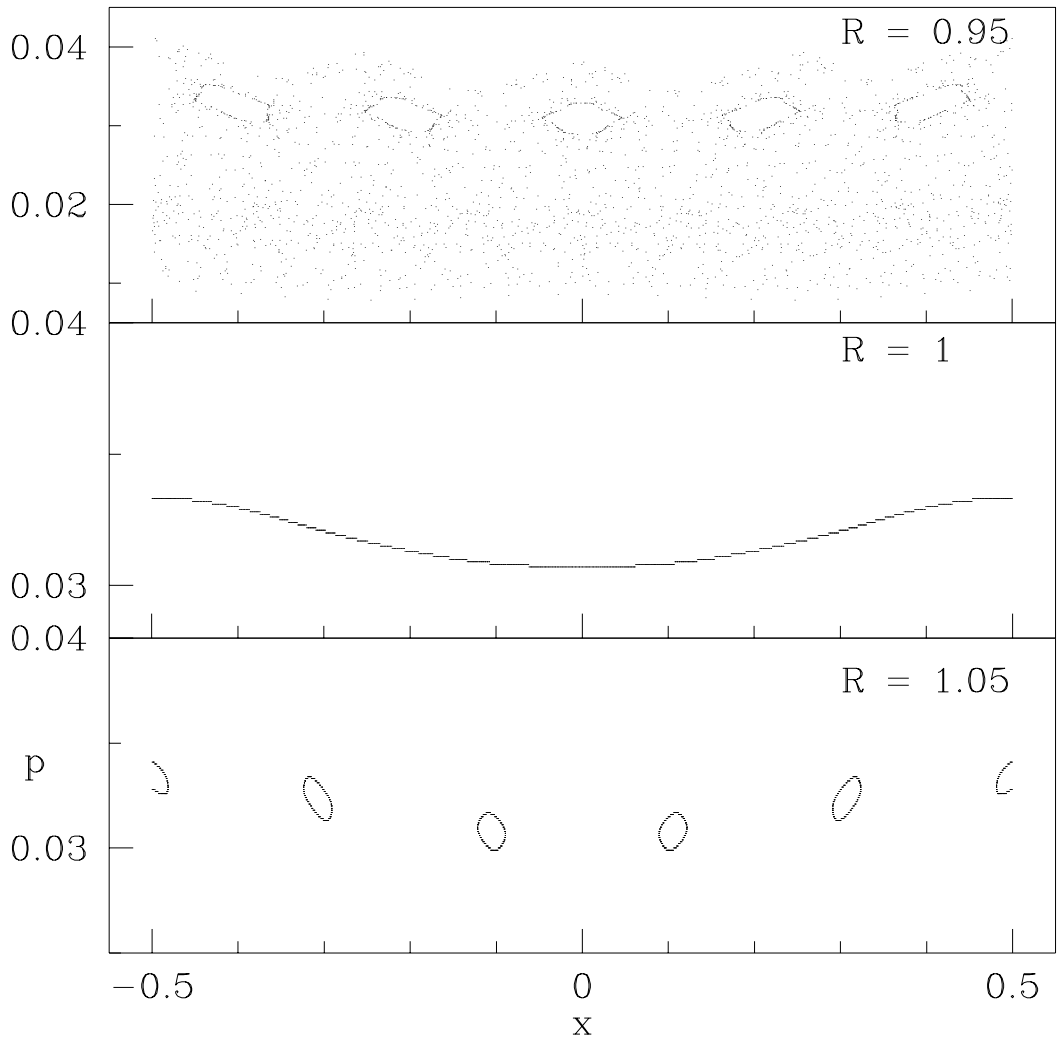,height=15cm,width=15cm}}
\caption{Shown are orbits of Fig. \ref{xp} having identical inital conditions. 
The initial condition corresponds to a KAM torus in the lower momentum region 
for $R=1$ (the negative momentum region is not shown here). Note the abrupt 
change in the stability and the non-generic features of the resultant phase 
space structures.}
\label{kam}
\end{figure}

Under the perturbation we can expect roughly that states whose absolute value 
of the initial momentum is less than $\sqrt{2 |\epsilon|}$ will be most 
affected. Thus low energy states will be most affected by the time-dependent 
forces. Fig. \ref{xp} shows the effect of the parameter $R$. While for $R=1$ 
the system is essentially KAM and has KAM tori interspersed with resonances, 
any small deviation of $R$ away from unity destroys low energy KAM curves and 
leads to increased chaos. Fig. \ref{kam} shows the fate of an individual KAM 
torus for which $R=1$ is a ``bifurcation'' point in parameter space and changes
stability on either side. We expect such behaviour to be generic to a large
class of similar systems and in this paper we will exhaustively study a
``standard map'' version of these systems. Just as the standard map provides 
an abstracted view of behaviour around nonlinear resonances we expect our 
model's analysis to provide such a view for these systems.

Many models have been studied where the time dependence $f(t)$ is a 
train of Dirac delta functions, the periodically kicked systems
\cite{reichl}. The reasons are evident: they are the
simplest Hamiltonian systems where many generic features of chaotic
complex systems may be observed; and they allow a partial integration
of the equations of motion, from kick to kick, enabling us to study
iterative ``mappings'' rather than differential equations. Quantum
mechanically the simplification enables us to partially integrate the
Schr\"{o}dinger equation and write the kick to kick propagator, or
Floquet operator, analytically \cite{berry79}.

An important paradigm in this class is the delta kicked rotor from which is 
derived the standard map \cite{chirikov79,chirikov87}. The dynamical richness 
of the classical system  which obeys the KAM theorem and the consequent smooth 
transition to chaos is now well known, and fairly understood 
\cite{greene79,shenker82,lich}. The corresponding quantum system has also 
been studied quite extensively as a model of ``quantum chaos'' (see
\cite{izrailev90} for an early review on this), and continues to provide an 
excellent model to numerically test our understanding of such 
systems \cite{arul99}.

The periodic input of energy into the system through the kicks can
result in a diffusive increase of the momentum, and strong 
kicking strength can lead to unbounded energy growth classically.
However, an important result on quantization is that the
eigenfunctions (quasienergy states) are generically {\it exponentially
localized in momentum space}, which  suppresses the momentum diffusion
even in the highly chaotic regime. A plausible mechanism for this
``dynamical localization'' was suggested when an analogy was found
\cite{grempel84} to Anderson type localization of electrons in
random on-site potentials \cite{anderson58}. Experimental realizations
of the delta kicked rotor, with cold atoms in pulsed standing
laser fields \cite{moore95}, has confirmed the quantum suppression of
diffusion. The localized states in unbounded momentum space results in
quasi-independence of the quasienergies, and the random matrix
properties \cite{brody81,haake} expected of quantized chaotic systems
are not seen. For instance the nearest neighbour spacing distribution
is Poisson rather than Wigner and the eigenfunction components are
not gaussian distributed. It must be stated that most of these results
are numerical and larger matrix calculations that are ``more
semiclassical'' may show spectral transitions as the bulk of the
eigenfunctions spread out and overlap with each other while the tails
are still exponentially localized. Some evidence of this will also be
present in this paper.

Following our motivational discussion above, we replace the time
dependence by a series of delta functions to facilitate the derivation
of a map in which we can study non-KAM behaviour of the kind suggested
above. Recently study of such systems has begun
\cite{bambi99,sankar01}. For instance in \cite{bambi99} 
it is shown that the quantum states are extended and delocalized in
the highly chaotic (strong field) regime.  In turn, the spacing
distribution of the quasienergies, unlike the kicked rotor, follow the
Wigner distribution. It is argued that the extended states do overlap
and hence the corresponding quasienergies are not independent,
resulting in level repulsion. Part of the present paper also
critically examines these results. In \cite{sankar01} a classical
analysis of a generalized system has been carried out to understand
the changes of stability that occur as a function of $R$ and some of
these results will be summarized below.

\section {Classical System}

The system of interest is a particle inside the potential $V_{sq}(x;a)$ 
in the presence of a particular time-periodic impulse. We consider the 
Hamiltonian given by

\begin{equation}
H = H_0 + \epsilon \cos\left({2\pi x/\lambda}\right)
\sum_{n=-\infty }^{\infty }\delta (n-{t/T}).
\label{ham}
\end{equation}

\n Kick-to-kick dynamics of the particle immediately after each pulse can
be described by an area preserving map which in dimensionless form is

\begin{equation}
\begin{array}{lll}
X_{n+1} &=& {(-1)}^{B_n} \left\{(X_n + P_n) - 
\mbox{Sgn}(P_n) B_n \right\} \\[6pt]
P_{n+1} &=& {(-1)}^{B_n} P_n + (K/2\pi ) \sin(2 \pi RX_{n+1}).
\end{array}
\label{cmap}
\end{equation}

\n Here $B_n=\left[\mbox{Sgn}(P_n)(X_n+P_n)+ 1/2\right]$ is the number 
of bounces of the particle between the walls during the interval between 
$n$th and $(n+1)$th kick, $[...]$ stands for the integer part of the argument. 
The state of the particle just after the $n$th kick is now given in the new 
variables as $X_n,P_n$. The sign of the momentum ($\pm 1$) is given by 
$\hbox{Sgn}(P_n)$. The following scaling relations are used to redefine the 
variables and parameters:

\begin{equation}
X_n = {x_n \over 2a}, \;\;\;\;\;
P_n = {p_nT \over 2a}, \;\;\;\;\;
K = {2\epsilon{\pi}^2 T^2 \over a \lambda}, \;\;\;\;\;
R = {2a \over \lambda}. 
\label{scaling}
\end{equation}

We note that $|X_n| \le 1/2$ and effective parameters of the particle dynamics
are $K$, the field strength and $R$, the ratio of the two length scales.
The map in (\ref{cmap}) shows the principal features that we have discussed 
earlier in the introduction and may qualify as a ``standard'' map for such 
non-KAM systems. In fact there is a close relationship between this 
``well-map'' (\ref{cmap}) and the standard map itself that allows us to study 
it as a {\it Generalized Standard Map} (GSM) \cite{sankar01}. This is the map:

\begin{equation}
\begin{array}{lll}
P_{n+1} &=& P_n + (K/2\pi )  \sin(2\pi RX_n) \\[6pt]
X_{n+1} &=& X_{n} + P_{n+1} 
\hspace*{0.6cm} (\mbox{mod}\;\;1)
\end{array}
\label{gsm}
\end{equation}

\n which is defined on a cylinder $(-\infty,\infty) \times [-1/2,1/2)$.
The well-map and the GSM differ only by boundary
conditions i.e., the former and latter have reflective and periodic
boundary conditions respectively. It is easy to see that this
does not play any effective role, in the sense that the trajectories
as evolved under the two systems can at most differ by a sign,
depending on the number of bounces undergone. 
This fact simplifies considerably our analysis of the well-map. 
When $R=1$, the GSM is the well studied and fairly understood standard 
map of the delta-kicked rotor.

\begin{figure}
\centerline{\psfig{figure=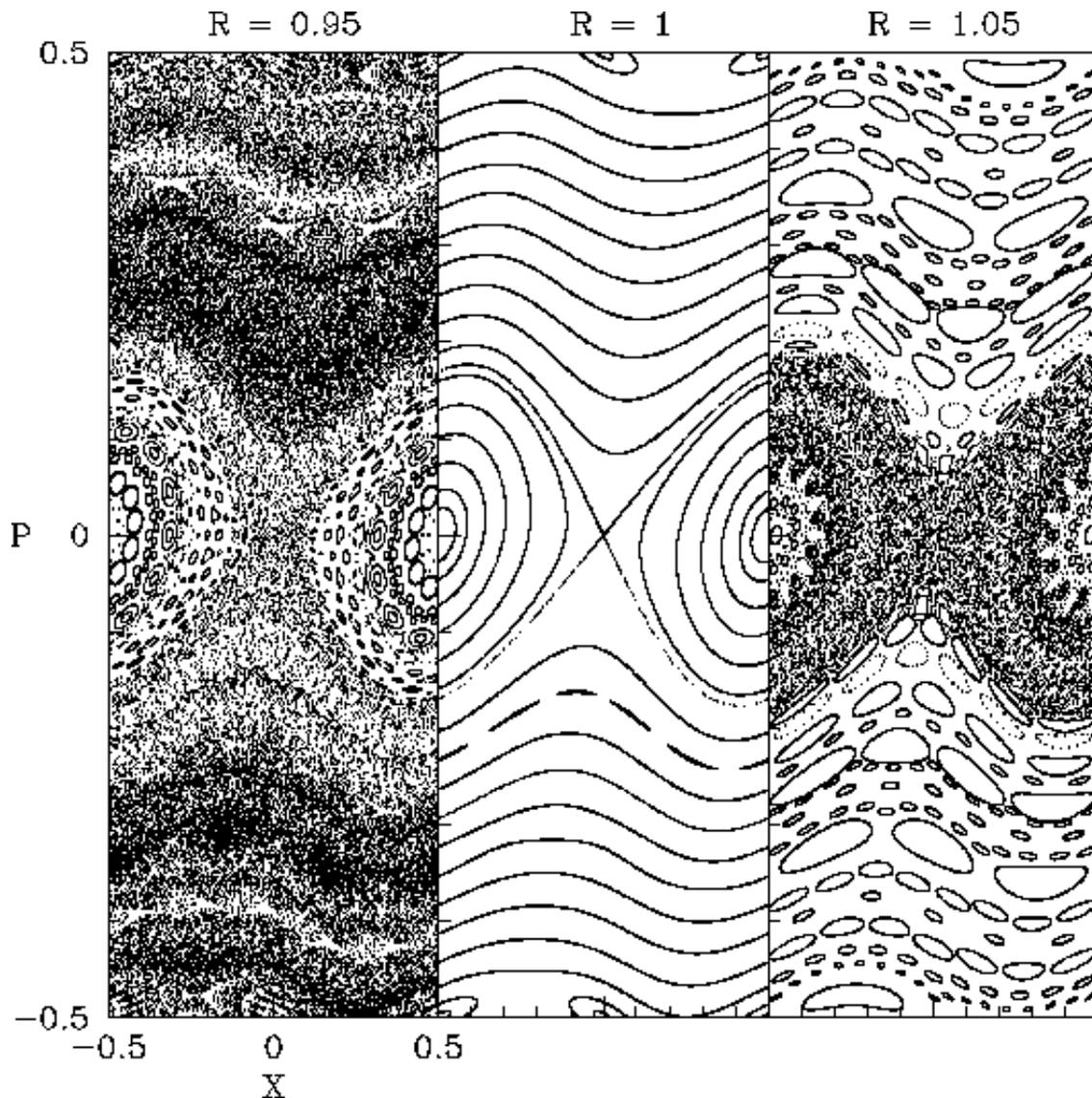,height=15cm,width=15cm}}
\caption{Phase space portrait of the GSM with $K=0.3$. For $R=1$, the dynamics 
is nearly regular wherein many smooth KAM tori are seen. When $R$ departs from
unity the dynamics is increasingly complex and no KAM tori are seen. This may 
be compared to the lower momentum region in Fig. \ref{xp}.}
\label{xp1}
\end{figure}

Dynamics of the GSM is highly chaotic and diffusive in the strong
field regime ($K \gg 1)$. In addition, it exhibits other interesting
features like the development of chaos and hence diffusion even in the
weak field regime ($K<1$) when $R\neq j$ where $j$ is a positive
integer, see Fig. \ref{xp1}. Note that the phase space portrait of 
both the well-map and the GSM are same.
Such dynamical features are in fact common 
in non-KAM systems and one such situation is shown earlier. We can understand
the development of chaos at low field strengths from the observation 
that the otherwise continuous map is discontinuous when $R\neq j$.
The KAM theorem does not hold for the discontinuous case and no smooth KAM
tori exist in the phase space, however small $K$ may be.  In the
absence of KAM tori, the phase space is chaotic and diffusive even in the
weak field regime. More over, when $R<1/2$ the GSM is a hyperbolic 
system. For a more detailed investigation of the GSM we refer the reader to
\cite{sankar01}.

\section {Quantum mechanics of the trapped particle}

For the Hamiltonian which is periodic in time with period $T(=2\pi/\omega)$, 
solutions of the Schr\"{o}dinger equation satisfy the eigenvalue 
equation \cite{zeldovich}
\begin{equation}
U|\psi_j\rangle = e^{-i\alpha_jT/\hbar}|\psi_j\rangle
\label{eigen}
\end{equation}

\n where $U$ is the one period time evolution operator. Here 
$|\psi_j\rangle$ and $\alpha_j$ are the quasienergy states and quasienergies
respectively. It is to be noted that the inner product of two arbitrary 
solutions of the Schr\"{o}dinger equation for arbitrary time dependent 
Hamiltonian is independent of time due to hermiticity of the Hamiltonian. 
As a consequence of this, states correspond to $\alpha_i$ and $\alpha_j$ such 
that $\alpha_i-\alpha_j\neq n\hbar\omega$ (where $n$ is an integer) are 
orthogonal at a given time. If $\alpha_i-\alpha_j=n\hbar\omega$, the states are 
degenerate and hence the quasienergies are uniquely defined with modulo 
$\hbar\omega$. From the set of all orthogonal states we may write the general 
solution at a given time as $|\Psi\rangle = \sum_j c_j|\psi_j\rangle$.

\subsection {Matrix Representation of U}

Periodically kicked systems are particularly easy to study since 
$U$ can be written immediately by integrating the Schr\"{o}dinger equation 
between successive kicks. For the Hamiltonian $H$ in Eq. (\ref{ham}) 
we have 

\begin{equation}
U = \exp\left\{-ik\cos\left({2\pi x\over\lambda}\right)\right\} 
\exp\left\{-i{H_0 T\over\hbar}\right\}
\end{equation}

\n where $k=\epsilon T/\hbar$. Note that the above time evolution operator 
is the quantum counter part of the well-map and not that of the GSM. The 
eigensystem in Eq. (\ref{eigen}) may be solved by
diagonalizing a matrix representation of $U$. The natural choice of basis
for the $U$-matrix is the eigenstates of the unperturbed Hamiltonian $H_0$:

\begin{equation}
H_0|n\rangle = E_n|n\rangle 
\end{equation}

\n where $n=1,2,3, \ldots.$ The energy eigenfunctions and eigenvalues are 

\begin{equation}
\langle x|n\rangle = \left\{\begin{array}{ll}
         {1 \over \sqrt{a}} \cos({n\pi x \over 2a}), & \mbox{for $n$ odd}\\
         {1 \over \sqrt{a}} \sin({n\pi x \over 2a}), & \mbox{for $n$ even}\\
                \end{array} \right. \;\;\; ; \;\;\; 
E_n = {n^2{\pi}^2{\hbar}^2 \over 8a^2}.
\end{equation}

The unitary matrix $U$ is then calculated as:  

\begin{equation}
U_{mn} = \langle m|U|n\rangle = \langle m|\exp\{-ik\cos(2\pi x/\lambda)\}
|n\rangle e^{-in^2\tau} \equiv F_{mn}e^{-in^2\tau}
\label{umn}
\end{equation}

\n where we have defined an effective Planck constant as

\begin{equation}
\tau = {{\pi}^2\hbar T\over 8a^2}
\label{tau}.
\end{equation}

\n As the external field preserves parity we have  

\begin{equation}
F_{mn} = \left\{\begin{array}{ll}
	0, & \mbox{if $m+n$ is odd}\\
	{1 \over 2\pi} \left\{Q_{m-n\over 2} - 
	{(-1)}^n Q_{m+n\over 2}\right\}, 
	   & \mbox{if $m+n$ is even}\\
        \end{array} \right. 
\label{F_matrix}
\end{equation} 

\n where 

\begin{equation}
Q_l = \int_{-\pi}^{\pi} \cos(l \theta)\ e^{-ik\cos(R\theta)}\ d\theta
\label{int}
\end{equation}

\n and $\theta =\pi x/a$. We note that $Q_l$ is a Bessel function integral
for integer $R$, while for non-integer $R$ the integral constitutes a kind 
of ``incomplete'' Bessel function. Invoking the Bessel function $J_s(k)$
through the following identity 

\[ e^{-ik\cos\theta} = \sum_{s=-\infty}^{\infty}{(-i)}^s
J_s(k)\ e^{-is\theta} \]

\n the integral can be evaluated as a series:
\begin{equation}
Q_l = 2\pi J_0(k) \ \delta_{l,0} + 2\sum_{s=1}^{\infty}{(-i)}^sJ_s(k) \ C_s
\label{int_exp}
\end{equation}

\n where 

\[ C_s = \int_{-\pi}^{\pi}\cos(l \theta)\cos(sR\theta) \ d\theta =
\left\{\begin{array}{ll}
{{(-1)}^l2sR\sin(sR\pi) \over {(sR)}^2-l^2}, & \mbox{for $sR \neq |l|$}\\
\pi, & \mbox{for $sR=|l|$}\\
                \end{array} \right. \ . \]

\n The relation $J_{-s}(k) = {(-1)}^sJ_s(k)$ has been used in 
Eq.(\ref{int_exp}). Note that if R is an integer, Eq.(\ref{int_exp}) 
simplifies to

\begin{equation}
Q_l = 2\pi \sum_{s=0}^{\infty} {(-i)}^{s}J_{s}(k) \ \delta_{|l|,sR}
\label{int_exp1}
\end{equation}

\n and a single term is picked out of the infinite series.

The forms of $Q_l$ allow us to assess the fall of the matrix elements of the 
unitary matrix $U$. For integer $R$ the unitary matrix can be essentially 
banded as the matrix elements fall off exponentially after a certain cut-off. 
For $R=1$, as is well known and can be seen from above for $l > k$ the matrix 
elements fall off exponentially, where $l$ measures the distance from the 
diagonal. On the other hand when $R$ is not an integer, apart from the Bessel 
function terms there are terms that are falling only polynomially in $l$. For 
instance when $R=1/2$ we have

\begin{equation}
Q_l = 2\pi{(-1)}^{l}J_{2l}(k) + {(-1)}^l 8\sum_{s=1,3,5...}^{\infty}
{{(-i)}^s s \sin (s\pi/2) \over s^2 - 4l^2} J_s(k) .
\end{equation}

\n The infinite series gets effectively cut-off for $s>k$. The finite
sum has terms that only decay as $l^{-2}$. Thus non-integer $R$ values
imply an important characteristic of the unitary quantum map: the
polynomial fall of matrix elements, as opposed to the exponential fall
characterizing integer $R$. In fact we may speculate whether non-KAM
systems are {\it always} characterized by polynomially decaying matrix
elements in the unperturbed basis. According to earlier studies 
eigenfunction localization crucially depends on the way in which 
matrix elements fall. 

When $R=1$, the classical equivalence of the kicked rotor to the
particle in a well was noted above. It is easy  to see that 
quantum mechanically also the equivalence persists. The parity symmetry
reduced rotor unitary matrix is identical to the well unitary operator
in this case and hence odd states of the rotor correspond to the odd
states of well, while the even states have a similar relationship.
Thus all that is known for the quantum standard map, including
exponential localization of eigenstates, may be carried over to the
well system with $R=1$. This allows us to address interesting 
questions of deviations from the standard map in a single model. 

The perturbing potential $\cos(2\pi x/\lambda )$ 
preserves the parity of $H_0$, and hence $U$ has the symmetry of
parity. In what follows we consider only the states which have odd
parity. In addition, the system has a spatial translational symmetry
when $R$ is an integer. Let us define a transformation for integer $R$ as

\begin{equation}
{\cal T} f(X) =f((X + 1/R)\,\mbox{mod}\,1)
\end{equation}

\n such that ${\cal T}^Rf(X)=f(X)$. ${\cal T}$ has the eigenvalues 
${\beta}_l=\exp(i{2l\pi\over R})$ where $l=0,1,2...(R-1)$. The commutation 
relation $[U,{\cal T}]=0$ leading to ${\cal T}|\psi\rangle = \beta_l
|\psi\rangle$. For $R=2$, $\beta_l = \pm 1$; in this case we consider only 
the states which correspond to $\beta_l=1$.

Dimensionless quantum parameters $k$ and $\tau$ are related to the classical 
parameters through the relation $K/R = 8k\tau$. The semiclassical limit 
is $k \rightarrow \infty$ and $\tau \rightarrow 0$, such that $k \tau$ is 
fixed. Any arbitrary state of the system at a given time is $|\Psi(t)\rangle =
\sum_n A_n(t)|n\rangle$ and its time evolution is given by
$A_m(t+T) = \sum_{n}U_{mn}A_n(t)$. 

\subsection {Quantum Resonance}

Here we investigate if the parameter $R$ has any effect on the important
phenomenon of ``quantum resonance''. We notice that the unperturbed motion 
of the particle, given by the Hamiltonian $H_0$, between the kicks simply 
adds phase to the wave function components (when expressed in the unperturbed 
basis, as in Eq. (\ref{umn})). At resonance ($\tau = 2\pi$), the unperturbed
motion between the kicks is absent. In this case, without loss of generality, 
the time evolution of an arbitrary state of the system is 

\begin{equation}
|\Psi(t)\rangle = e^{-ik\cos(2\pi x/\lambda)t}|\Psi(0)\rangle
\label{qres}
\end{equation}

\n and thus $|\Psi(t)|^2 = |\Psi(0)|^2$. Note that here $t$ is the number of
kicks. The kinetic energy of the particle is then

\begin{eqnarray}
E(t) = E(0) &+& {-{\hbar}^2 \over 2}
\left\{ \left({4\pi kt \over \lambda}\right)
\int_{-a}^{a} \sin\left({2\pi x \over \lambda}\right) 
\mbox{Re}\left\{i{\Psi}^{*}(0)
\frac{\partial \Psi(0)}{\partial x}\right\}dx \right. \nonumber \\ 
&-& \left. {\left({2\pi kt \over \lambda}\right)}^2 
\int_{-a}^{a} |\Psi(0)|^2 \sin^2\left({2\pi x \over \lambda}\right)dx
\right\} . 
\end{eqnarray}

\n In the limit $t \rightarrow \infty$ energy grows quadratically with the
number of kicks. If $|\Psi(0)\rangle = |n\rangle$, i.e., the initial state
is one of the unperturbed state itself, the energy is purely quadratic. 
In fact, the energy can be found exactly as

\begin{equation}
E(t) = E(0) \left\{1+{\left({ktR \over n}\right) }^2 (2-A) \right\}
\end{equation}

\n where
\[ A=\left\{\begin{array}{ll}
{\sin(2\pi R) \over \pi R} \left({n^2 \over n^2 - 4R^2}\right) 
& \mbox{if $n \neq 2R$}\\
{(-1)}^{n+1} & \mbox{if $n=2R$}
                \end{array} \right. \ . \] 

Since $A\neq 2$, we observe that the quadratic energy growth is unaffected
by the length scale ratio $R$. Numerically we have found that 
this behaviour is seen when $\tau$ is rational multiples of $2\pi$ also. 
Thus the quantum resonance phenomena of the well system is very similar to
that of the kicked rotor \cite{chirikov79,izrailev80}. It is to be noted that
resonance is a non-generic pure quantum phenomena and no correspondence to it
can be seen in the classical system. In the context of a particle in a well, 
quantum resonance may lead to enhanced ionization in a finite well.

\section {Results}

Having given sufficient description of the system under investigation, here we 
analyze quasienergy states and quasienergies of the generic quantum system 
($\tau$ is irrational multiples of $2\pi$) in relevant classical regimes. On 
taking a truncated $N$ dimensional Hilbert space spanned by the first $N$ 
unperturbed basis states which belong to odd parity, diagonalization of the 
matrix $U_{mn}$ gives the eigenstates $\{|\psi\rangle\}$ such that 
$|\psi\rangle = \sum_n \psi_n|n\rangle$. We consider only states that are 
``converged'' in the sense that they are independent of the truncation size 
$N$. Thus the states we are interested in belong to that of the infinite 
Hilbert space; they are states of the infinite cylinder and {\it not} of a 
truncated cylinder, or torus. The last distinction becomes important as 
quantum states that belong to the cylinder can have completely different 
localization features than those that belong to a truncated cylinder. As we 
are interested in a particle in an infinite potential well, such a truncation 
lacks physical meaning.
 
\subsection{Localization measures of eigenstates}

Localization can be measured using a unified quantity, the Renyi Participation 
Ratio $\xi_q$ 

\begin{equation}
\xi_q=\left(\sum_{n}|\psi_n|^{2q}\right)^{1/(q-1)}
\end{equation}

\n of which the entropy and participation ratio (PR) are special cases. 
In our analysis we first use a normalized information entropy as a measure
of localization of states, and this is defined as:

\begin{equation}
S = {-1\over \ln{(N/2)}}\sum_{n=1}^{N}{|\psi_n|}^2\ln {|\psi_n|}^2.
\end{equation}

\n It is easy to see that $S=\ln\xi_1/\ln(N/2)$. This measure compares the 
entropy to that of the eigenfunctions of $N \times N $ matrices belonging 
to the Gaussian Orthogonal Ensemble (GOE) which is approximately $\ln(N/2)$.
The GOE is relevant to time reversal symmetric systems such as we are 
considering.  

\begin{figure}
\centerline{\psfig{figure=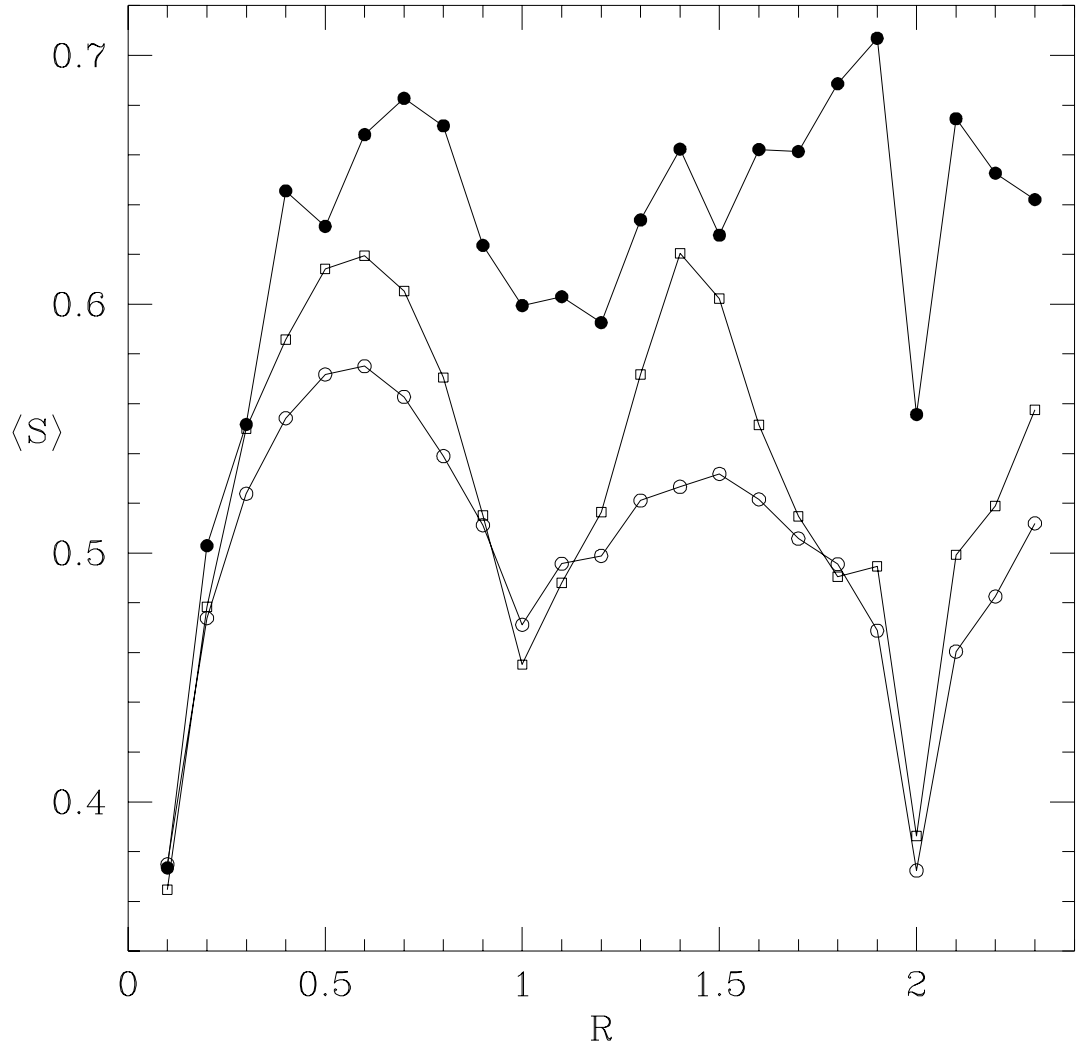,height=15cm,width=15cm}}
\caption{Average entropy of 1000 eigenstates for : $K = 0.1,\tau = 0.001$ 
($\circ$); $K = 1,\tau = 0.01$ ($\Box$); $K = 10, \tau = 0.1$ ($\bullet$). 
$N=1200$ in all the cases.}
\label{avg_ent}
\end{figure}

First we calculate a gross measure of localization in a given spectrum by 
averaging over all converged states. We set criteria for the states to be 
converged so that the states belong to the cylinder, or are at least very 
close to states that belong to the cylinder. In all the following cases, 
the eigenvalues are converged in modulus to unity to within 0.0001 or better. 
Fig. \ref{avg_ent} shows the average entropy as a function of $R$. For small 
$K$ ($\le 1$), the oscillations are qualitatively similar with 
distinct entropy minima at integer $R$ and maxima at around half-integer $R$. 
This may provide a simple mechanism to experimentally control the extent of 
localization. The information entropy is of course basis dependent; the 
unperturbed basis we use is a useful one as it has information about 
localization in the momentum.

Naturally, the minima in entropy is expected to have strong associations with 
the presence of stable regions in the classical phase space. Of special 
significance are KAM tori in phase space, as these structures are complete 
barriers to classical diffusion in momentum. In spite of the fact that in the 
classical system all the KAM tori break up in the standard map ($R=1$) at 
$K=1$, we observe a minimum entropy. This is due to the presence of cantori
which are partial barriers for chaotic orbits and suppresses global
diffusion. For non-integer $R$ a complex phase space picture emerges
and has been discussed in \cite{sankar01}.  Maximum entropy around
half-integer $R$ is the classical parametric regime where the
discontinuity is maximum corresponding to the maximum chaos assisted
diffusion. 
 
For large $K$ ($=10$), oscillations in entropy are still present while there 
is apparently complete chaos for all relevant $R$ values. We can understand 
these oscillations as due to the strong correlation between the localization 
of eigenstates and classical diffusion coefficient. For $R<1/2$ the 
semiclassical parameter $k=K/(8R\tau)$ is large, yet there is increased
localization of states due to limited classical diffusion, presumably
due to the presence of cantori. For the kicked rotor the exponential
localization length was found to be proportional to the classical
diffusion coefficient \cite{chirikov81}. This was found by numerical
experiments and is supported by certain qualitative arguments.  We are
now in a position to examine the relationship between quantum
localization and classical diffusion in the context of the particle in
a well, wherein we have the freedom of another control parameter,
namely $R$, with which to vary the classical diffusion.

\begin{figure}
\centerline{\psfig{figure=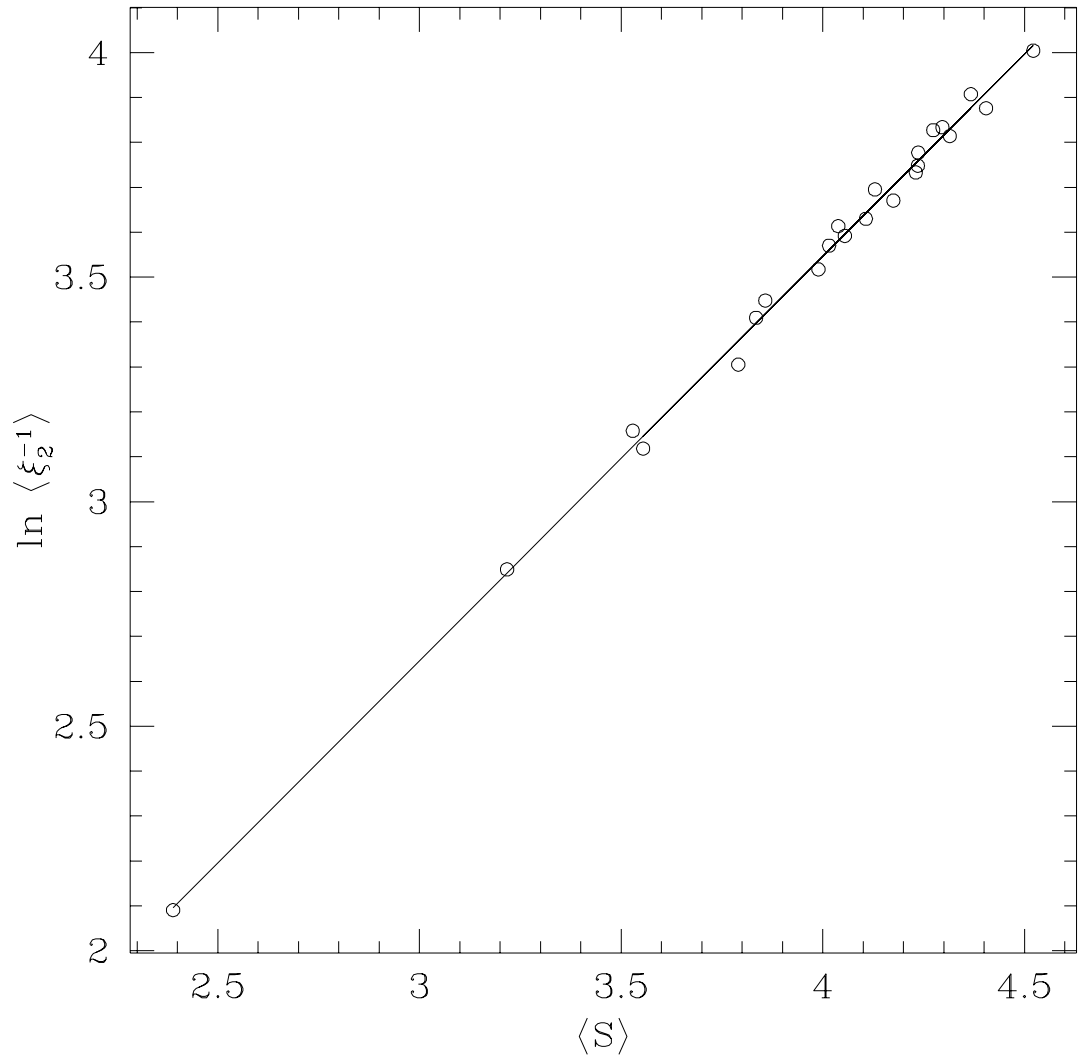,height=15cm,width=15cm}}
\caption{Average entropy .vs. the log of the  average PR corresponding to the 
case $K=10,\tau = 0.1$ of Fig. \ref{avg_ent}. The slope of the fitted 
straight line is $0.9\pm 0.01$.} 
\label{ep}
\end{figure}

Instead of studying localization lengths we study here the measures of 
localization such as the entropy or the PR. We study the PR more
closely rather than the entropy. In chaotic regimes we have numerically 
ascertained that exponential of the entropy is proportional to the PR, as 
shown in Fig. \ref{ep}. The relationship between the localization length 
hitherto calculated for the kicked rotor and the PR calculations we present 
will need more detailed study, but we expect them to be roughly proportional 
to each other. In fact if we assume a fully exponentially localized state with 
$|\psi_n| \sim \exp(-|n-n_0|/l_\infty)$, then the PR is

\begin{equation}
\label{partloc}
\xi_2^{-1}\,=\, (\sum_{n}|\psi_n|^4)^{-1} \,=\, 2 l_\infty.
\end{equation}

We recapitulate the argument connecting classical diffusion and the 
localization length for the specific system we are considering, as there 
are difference in factors. On considering time evolution of an initial state,
kinetic energy diffuses for a certain time $t_c$ and then attains 
quasiperiodic saturation. The number, $n_c$, of unperturbed states that are 
excited during the time evolution is related to the critical time by the 
diffusion equation:

\begin{equation}
\pi^2 \hbar^2 n_c^2=D_{\sc cl}t_c \;\;\; ; \;\;\;
\left \langle (p_t-p_0)^2 \right \rangle = D_{\sc cl}t
\end{equation}

\n where $D_{\sc cl}$ is the classical diffusion coefficient in momentum and
$\langle ..\rangle$ represents the ensemble average. Here the momenta and the 
diffusion coefficient have dimensions and we have taken $a=1/2$. The critical 
time being the Heisenberg time relevant for $n_c$ equally spaced eigenstates, 
we get $t_c \sim n_c T/2 \pi$. If the average localization length 
$\langle l_{\infty}\rangle$ is also $n_c$, we obtain the relation :

\begin{equation}
\langle \xi^{-1}_2 \rangle = 2\langle l_{\infty}\rangle = 
{\alpha\pi\over 4\tau^2}D(K,R)
\label{prd}
\end{equation}

\n where $\tau$ is the dimensionless effective Planck constant defined in
Eq. (\ref{tau}) and $\alpha$ is a constant whose value has been numerically 
determined as 1/2 for the standard map \cite{shep86}. $D(K,R)$ is the
dimensionless diffusion coefficient which one will get from using the
dimensionless maps Eq. (\ref{cmap}) or Eq. (\ref{gsm}). The dependence
on {\it both} $K$ and $R$ is emphasized.

\begin{figure}
\centerline{\psfig{figure=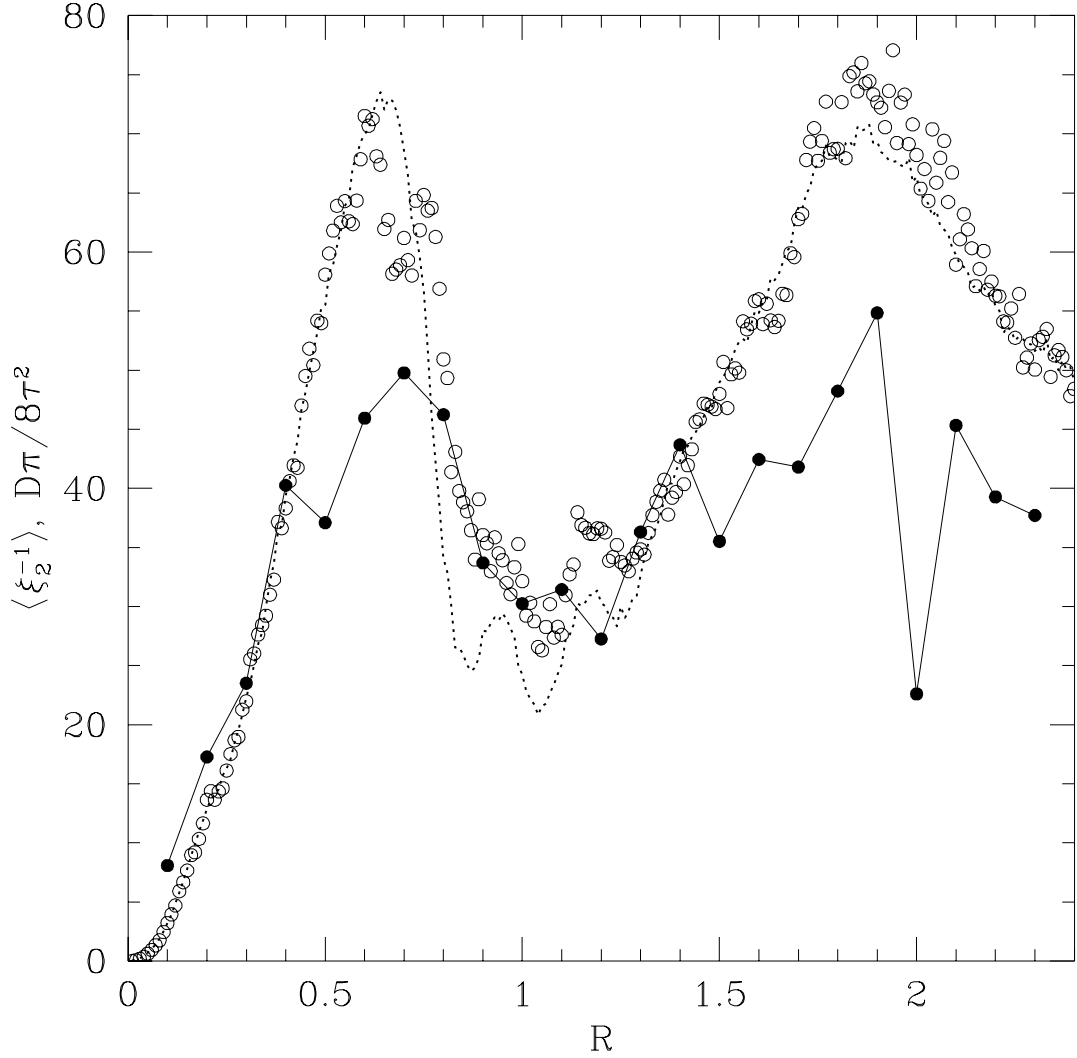,height=15cm,width=15cm}}
\caption{The average PR ($\bullet$) and the scaled classical diffusion 
coefficient ($\circ$) are plotted as a function of $R$ for the case 
$K=10,\tau = 0.1$. The dotted line is the scaled coefficient calculated 
using up to the second order time correlation. Higher order time correlations 
are insignificant since the classical system is highly chaotic.}
\label{pr_dcoef}
\end{figure}

\begin{figure}
\centerline{\psfig{figure=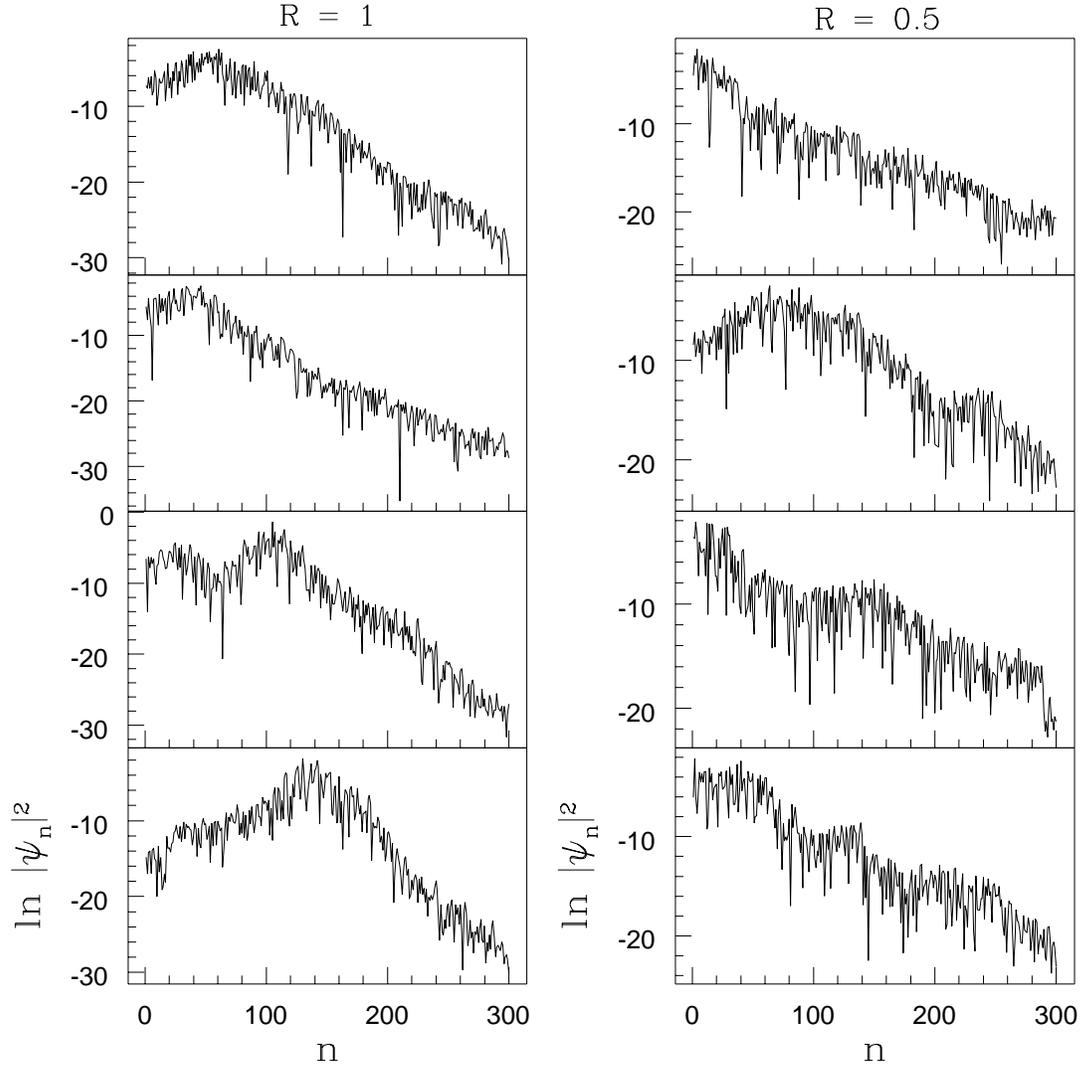,height=15cm,width=15cm}}
\caption{Typical eigenstates for the case $K=10,\tau = 0.1$. States 
corresponding to $R=0.5$ have more fluctuations compared to the rotor 
($R=1$) states.}
\label{evec_C10}
\end{figure}

In Fig. \ref{pr_dcoef} we show the average PR and the scaled diffusion 
coefficient according to the relation Eq.(\ref{prd}). We see that the 
relation derived above holds in some parameter regions while it picks up only
qualitative features of the oscillations in the others. In particular the 
relation seems to hold for $R<1/2$ when the classical system is hyperbolic 
as well as around $R=1$. The deviations from the relation (\ref{prd}) might 
be due to fluctuations of the state components in the unperturbed basis 
(one such case is shown in Fig. \ref{evec_C10}). These fluctuations may 
lead to different scaling behaviour between the average PR and the 
classical diffusion coefficient. However more detailed investigations are 
needed to make any positive statements. The sharp deviation for $R=2$ 
can be accounted for as due to the presence of an extra quantum symmetry 
discussed above.

\begin{figure}
\centerline{\psfig{figure=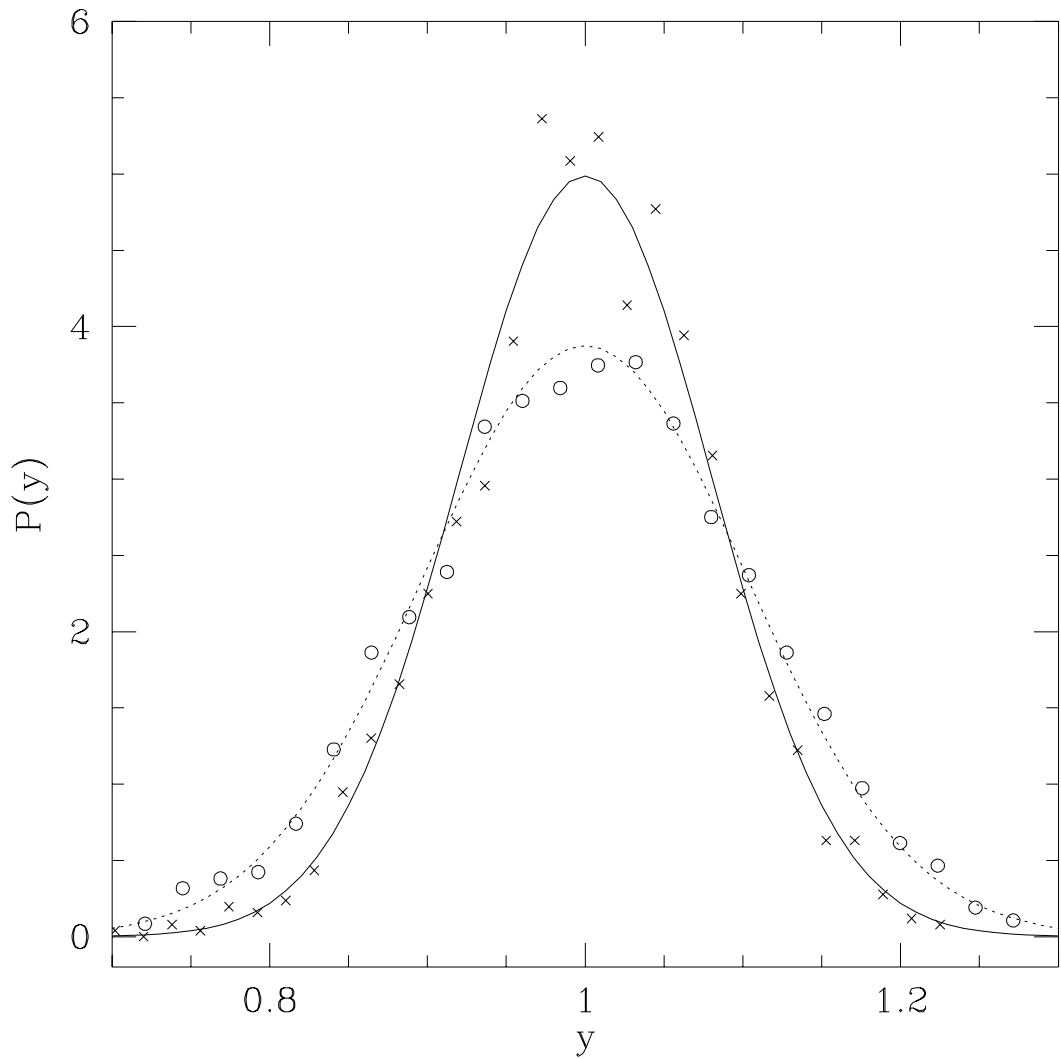,height=15cm,width=15cm}}
\caption{Probability distribution of $y$,  the normalized log of the PR, for 
the kicked rotor case ($R=1$) in the chaotic regime. Here we have taken 
$K=10$ and $\tau = 0.025$ ($\times$), $\tau = 0.05$ ($\circ$). Smooth curves 
are corresponding gaussian distributions.}
\label{dbn_lnpr1}
\end{figure}

Following our study on average PR and its scaling with the classical diffusion 
coefficient, we may then enquire about how the PR itself is distributed in a 
given spectrum if the average reflects the general behaviour. We find that when
the classical system is chaotic, the distribution of a normalized quantity 
$y= \ln\xi_2^{-1}/\langle \ln\xi_2^{-1} \rangle$ (this is similar to the 
distribution of the entropy due to the linear relationship exhibited above) 
is nearly normal as seen in Fig. \ref{dbn_lnpr1}. This may be attributed to 
a realization of the Central Limit Theorem. However the PRs and Inverse 
Participation Ratios (IPRs) themselves are not normal. Their distributions 
may be got by assuming that the distribution of $y$ is normal. Thus the PRs 
are distributed according to the lognormal distribution \cite{lognormal}:

\begin{equation}
\Lambda(\xi_2^{-1}) = {1\over \sqrt{2\pi}\sigma \langle \ln \xi_2^{-1} \rangle
\xi_2^{-1}} \exp\left\{-{1\over 2\sigma^2}\left({\ln\xi_2^{-1}\over
\langle \ln\xi_2^{-1} \rangle}-1\right)^2\right\}
\end{equation}

\n where $\sigma^2$ is the variance of $y$. As an immediate consequence,
distribution of IPRs is also lognormal. Distribution of such localization 
measures are much of significance. Recently the distribution of IPRs have been 
exploited to show that the distribution of resonance widths in wave-chaotic
dielectric cavities is lognormal \cite{narimanov00}.  

\begin{figure}
\centerline{\psfig{figure=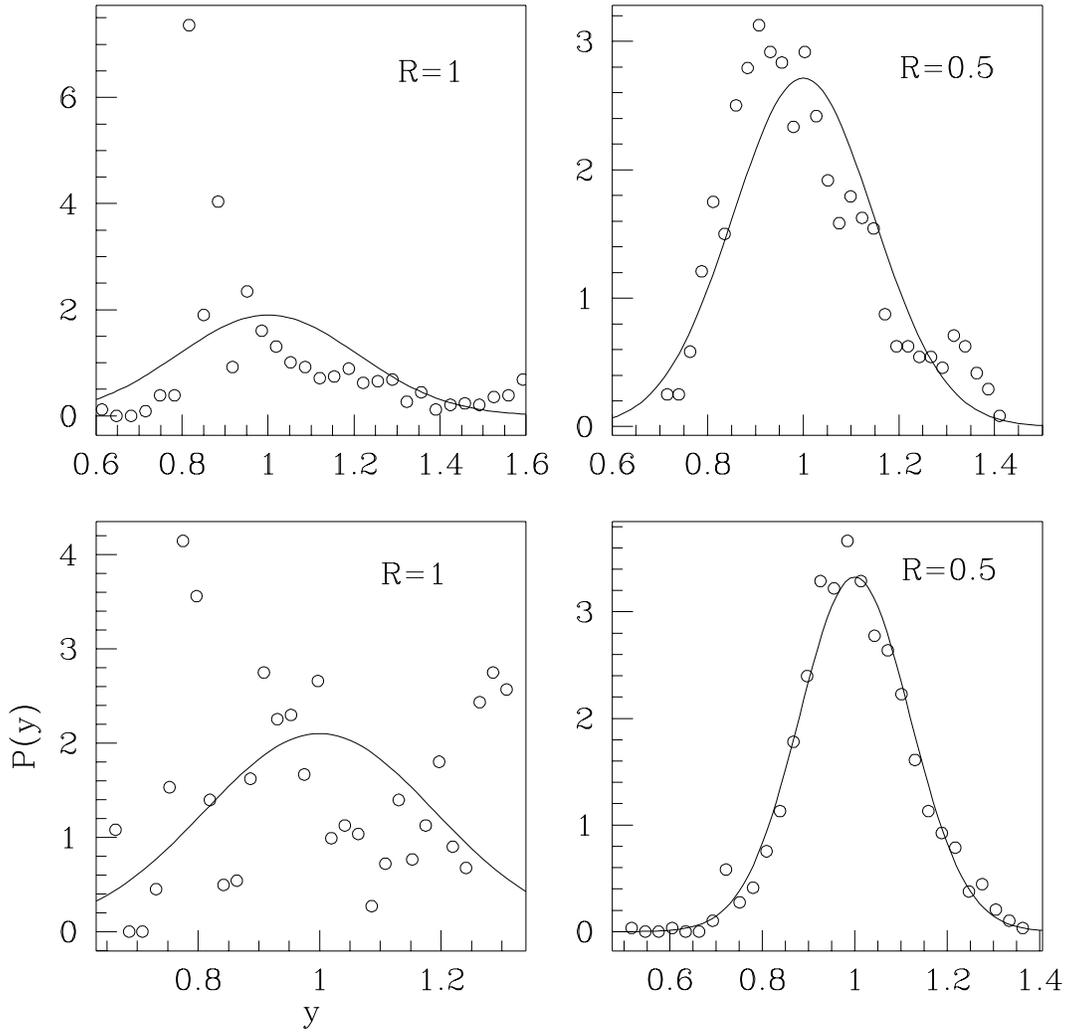,height=15cm,width=15cm}}
\caption{Probability distribution of $y$, the normalized log of the PR, for 
the case $K=0.1,\tau = 0.001$ (first row) and for $K=1,\tau = 0.01$ (second 
row). Smooth curves are  corresponding  gaussian distributions. Note the 
sensitivity of these distributions to the classical dynamics.} 
\label{dbn_lnpr}
\end{figure}

When $K$ is small ($\le 1$), the classical motion is nearly regular for $R=1$, 
while chaotic for $R \le 0.5$. However the time scales for classical diffusion 
is large making the observation of $R$ effects on quantum dynamics hard to 
discern. For instance the nearest neighbour spacing distribution may remain 
very close to the Poisson distribution. In such a situation we find that the 
distribution of the PRs provides a positive litmus test. In Fig. \ref{dbn_lnpr}
such an example is shown, wherein even for small field strengths the effect 
of $R$ is clearly visible as a tendency for $y$ to be normally distributed. 
This is an indication of the ``delocalization'' that is taking place in the
eigenfunctions. This delocalization is limited in the sense that while the 
eigenfunctions remain square integrable there is more spreading out in the 
bulk part of the states. Thus we may conclude that distribution of the
localization measures is a sensitive quantity in chaotic quantum systems.

\begin{figure}
\centerline{\psfig{figure=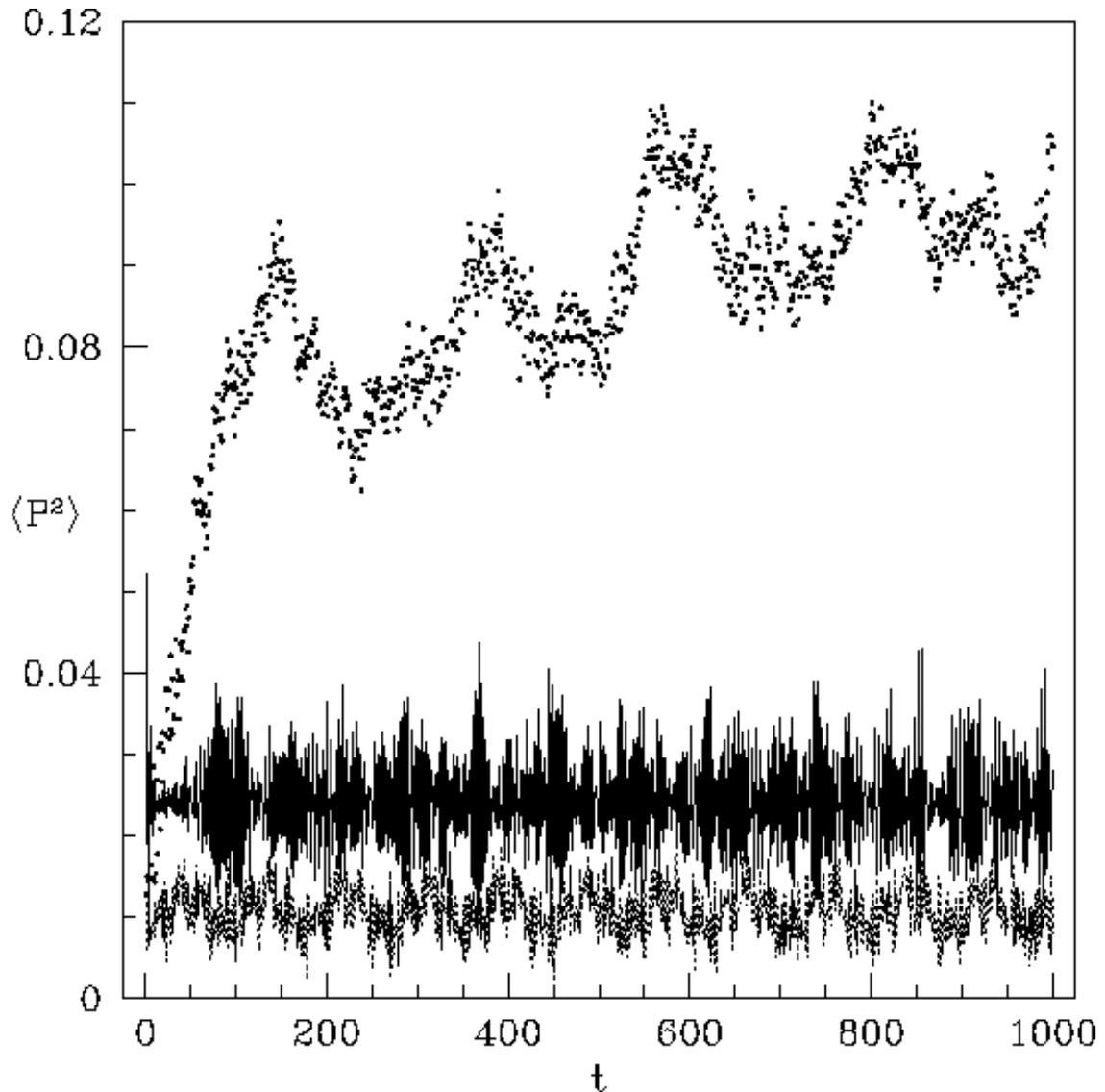,height=15cm,width=15cm}}
\caption{Shown is the scaled kinetic energy $\langle P^2\rangle$ of a state,
which is initially the ground state of the unperturbed system, as a function
of time. Here the parameters are $K=1,\tau = 0.01$ ; $R=1$ (solid line),
$R=1.5$ (dots) and $R=2$ (dotted line). Effect of non-integer $R$ is remarkably 
seen in the evolution as the kinetic energy of the quantum particle saturates 
at much higher value compared to integer $R$ cases.} 
\label{eng}
\end{figure}

Time evolution of non-stationary states must reflect the properties of the 
stationary states and is also of importance in the context of experiments. 
Here we have studied the diffusion in kinetic energy of a state $|\Psi\rangle$
which is initially the ground state of the unperturbed system. We illustrate
with one example wherein for a fixed classical parameter $K$, the effects of 
non-integer $R$ is seen clearly for a given $\tau$ value. Thus tuning $R$ 
essentially tunes $\lambda$ since $a$ is fixed through the relation 
(\ref{tau}). In Fig. \ref{eng} scaled kinetic energy $\langle P^2\rangle = 
\langle \Psi|P^2|\Psi\rangle$ is shown as a function of time (number of kicks) 
for a small value of $K$ corresponding to a small classical field strength 
$\epsilon$. We note that while the quantum diffusion saturates at a much 
higher value for $R=1.5$, compared to $R=1$, the actual classical field 
strength $\epsilon$ (from Eq. (\ref{scaling})) is {\it smaller} by a factor 
of $1.5$. For comparison is shown another integer case, $R=2$, wherein the 
classical diffusion is smaller compared to $R=1$. 

\subsection {Eigenvalues and Eigenstates}

It is clear from our earlier observations that the states are more localized 
in the regular or mixed regimes of the classical system while less localized 
(or delocalized) in the chaotic regimes. The degree of localization is also 
controlled by the ratio of the length scales and complexity of the classical 
phase space is reflected in the localization measures. Here we look at the 
quasienergies and the corresponding states more closely.

\begin{figure}
\centerline{\psfig{figure=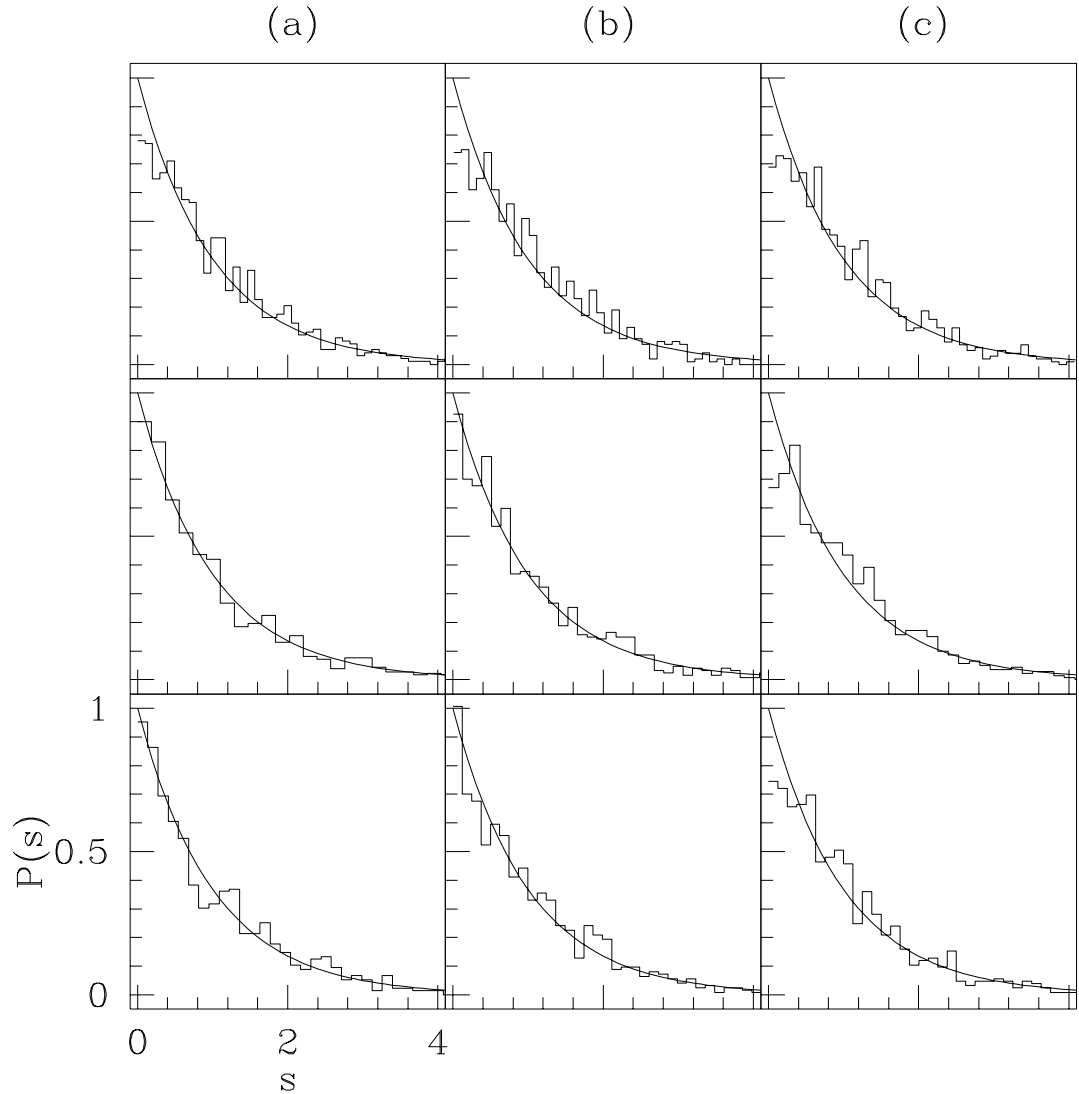,height=15cm,width=15cm}}
\caption{Nearest neighbour spacing distributions of 1000 quasienergies for 
(a) $K=0.1, \tau=0.001$; (b) $K=1,\tau=0.01$; (c) $K=10,\tau=0.1$ with 
$R=0.5,1,1.5$ (top to bottom) and $N=1200$. Smooth curves are Poisson 
distributions. Note the relative insensitivity of these distributions to the 
classical dynamics.} 
\label{nns1}
\end{figure}

In Fig. \ref{nns1} we have shown the nearest neighbour spacing
distribution of the quasienergies for various parameters. The first
row and the last column of the catalogue correspond to classically
chaotic regimes and the rest belong to the regular/mixed phase space
regimes. In regular/mixed regimes where the states are highly
localized, the spacing shows excellent agreement with the Poisson
distribution. On the other hand, in chaotic regimes the spacing
agrees well with the Poisson distribution except at small spacings. This is
due to the fact that bulk part of the eigenstates are delocalized
and they overlap each other.  However, the tail part of the states are
exponentially localized and the degree of overlap is not significant
enough. We also notice that the spacing distribution is only slightly
sensitive to the nature of the classical dynamics in the case of the
unbounded kicked rotor or the well, at least in the parameter regimes we have
investigated.  In such situations, as we have demonstrated earlier, the
distribution of PRs is a good measure to distinguish the chaotic
quantum systems from the regular systems.
 
Our extensive calculations of the eigenstates in chaotic regimes show that, 
in general, it is hard to qualitatively differentiate the states corresponding 
to non-integer $R$ values from the rotor ($R=1$) states as far as their 
localization behaviour is concerned. In particular it is not easy to 
distinguish the emergence of non-exponential tails unequivocally. However, we 
found that eigenstates corresponding to non-integer $R$ values generally have 
more fluctuations compared to the rotor states; this is illustrated with some 
examples in Fig. \ref{evec_C10}. The fluctuations are closer to the RMT 
predictions in the case of non-integer $R$ values and is shown further below. 

Recently there have been studies on the special case ($R=0.5$) of the
system (\ref{ham}), with the motivation of revealing quantal behaviour
of  non-KAM systems \cite{bambi99}. It was observed that the
quasienergy states are ``extended'' in the unperturbed basis and as a
result the spacing distribution was shown to be Wigner distributed. At
this juncture we would like to compare our results with certain
aspects of this work. In \cite{bambi99}, the eigenstate shown in the 
highly chaotic regime ($K=50,N=1024$; we have not been able to ascertain 
the value of $\tau$ used in this work) does {\it not} appear to belong to 
the unbounded phase space as it spreads all over the basis. Thus while states 
such as these may belong to some truncated dynamical system, it does not
belong to the infinite Hilbert space of the well system. Increasing
the dimensionality of the matrix used will modify such states; in
short they are not converged.  As we demonstrate below, unconverged or
poorly converged states may  mislead us in understanding the spectrum.

Large $K$ implies large $k$ for given $R$ and $\tau$, and hence our 
calculation demands bigger dimensionality $N$ of the truncated Hilbert space, 
since the PR is roughly increasing as $k^2$. Although we take $N=2000$, 
getting a good number of converged states is problematic. We pursue the 
spacing distribution with different convergence criteria for the numerically 
obtained states. The convergence criteria uses partial sum of the state 
components :

\begin{equation} 
{\{\hbox{Sum}\}}_M = \sum_{n=1}^{M} {|\psi_n|}^2 \;\;\; ; \;\;\; 
\hbox{with} \;\; M<N.
\end{equation}

\n For a well converged state we expect that 
${\{\hbox{Sum}\}}_M \approx 1$, even for $M\ll N$. We denote by $N_M$
the number of converged eigenstates whose ${\{\hbox{Sum}\}}_M$ is
greater than $S_M$ (an arbitrary number close to, but less than,
unity) for a fixed value of $M$. Thus the convergence criteria is
characterized by $M$ and $S_M$.

\begin{figure}
\centerline{\psfig{figure=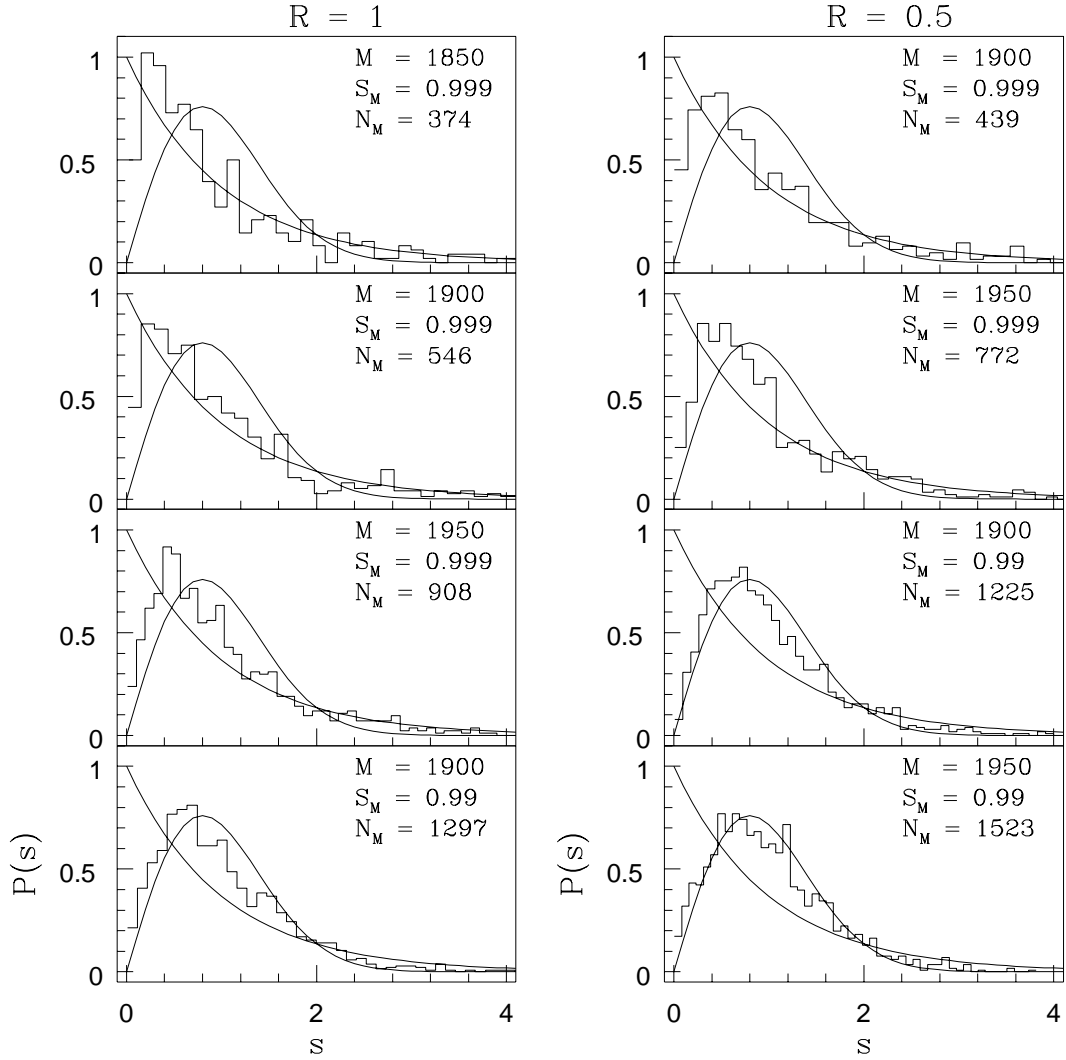,height=15cm,width=15cm}}
\caption{The nearest neighbour spacing distributions for the case $K = 50,
\tau = 0.1$. Smooth curves are the Poisson and Wigner distributions. 
Convergence criteria is relaxed as we move from top to bottom. A ``spectral 
transition'' is observed.} 
\label{nns2}
\end{figure}

In Fig. \ref{nns2}, we have shown the spacing distributions with different 
criteria for two cases. In both the cases transition to Wigner distribution is 
evident as the convergence criteria is relaxed. The unconverged or poorly 
converged states do not belong to the physical system of our interest and
the corresponding quasienergies follow the RMT prediction. Obviously, 
reliability of the result is more in the top plots where the spacing shows 
neither Poisson nor Wigner distributions. Though the tail part shows the 
Poisson behaviour there are significant discrepancies in the small spacing. 
A more correct picture may be closer to the scenario of the chaotic regimes 
presented in Fig. \ref{nns1}.  

\begin{figure}
\centerline{\psfig{figure=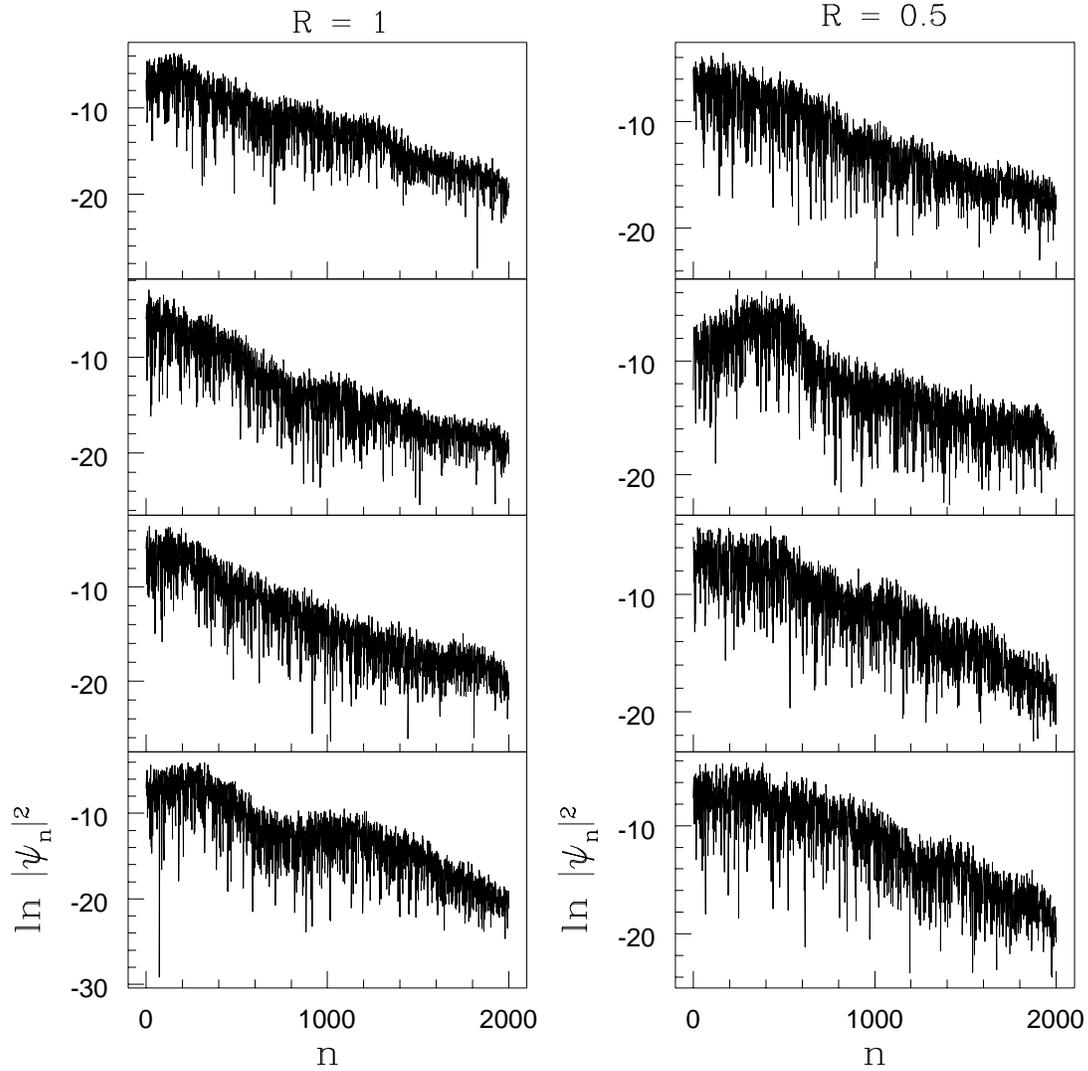,height=15cm,width=15cm}}
\caption{Typical well converged eigenstates for the highly chaotic case: 
$K=50,\tau=0.1$ and $N=2000$.}
\label{evec_C50}
\end{figure}

\begin{figure}
\centerline{\psfig{figure=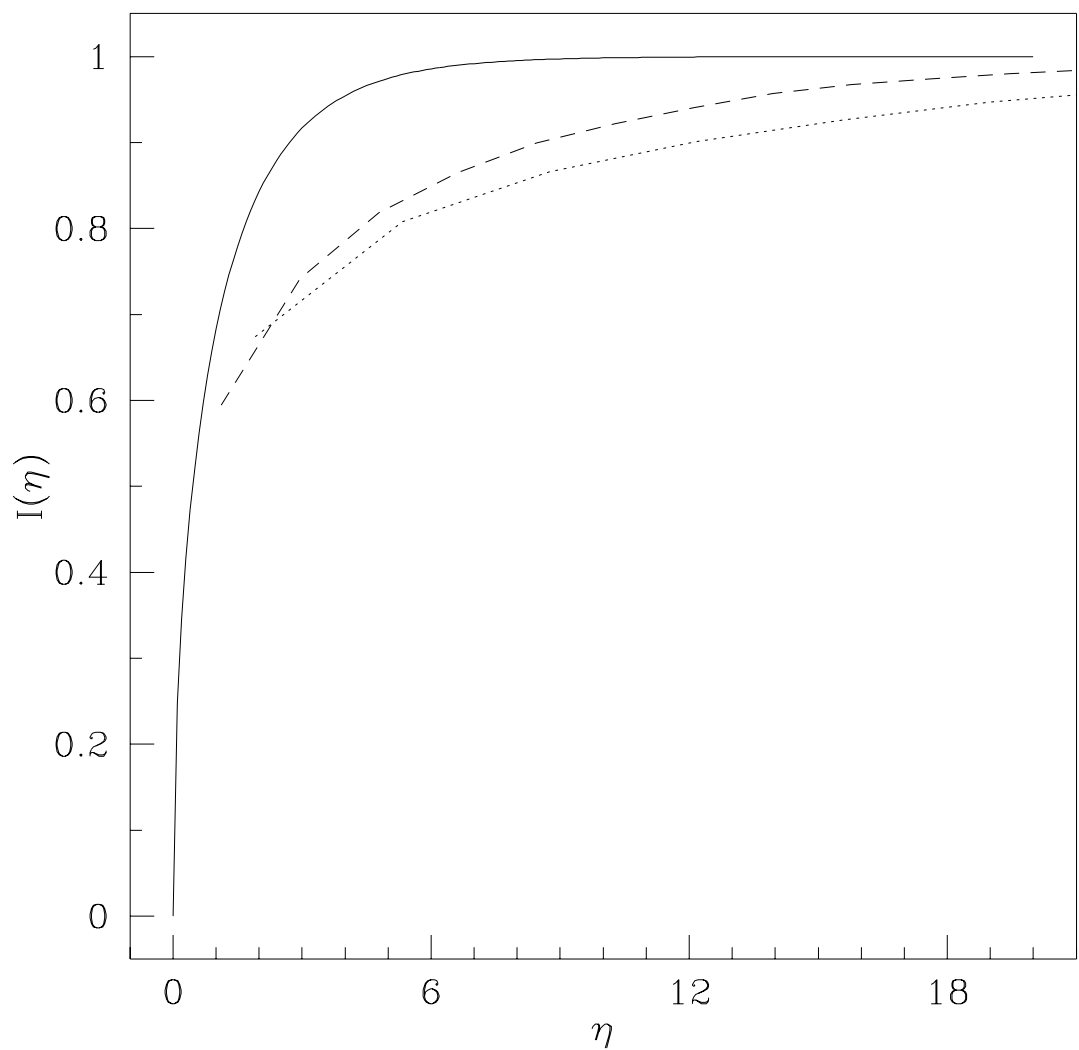,height=15cm,width=15cm}}
\caption{Collective cumulative distribution of the components of the states
shown in Fig. \ref{evec_C50}. Dotted curve corresponds to $R=1$ while dashed
curve corresponds to $R=0.5$. Solid curve is the cumulative Porter-Thomas
distribution.} 
\label{evdbn}
\end{figure}

Shown in Fig. \ref{evec_C50} are few ``well converged'' states, with more 
stringent convergence criteria ($M=1600, S_M=0.9999$). With this criteria we
have only $N_M=12$ and $4$ for $R=1$ and $0.5$ respectively. The state 
components exhibit strong fluctuations in the basis. Here again it is hard 
to differentiate the two cases qualitatively. The states corresponding to 
$R=0.5$ also appear to have exponential tails. To see the distribution of the
state components, we introduce a variable $\eta_n = |\psi_n|^2/\overline 
{|\psi_n|^2}$ where over bar stands for average over the state components 
such that $\overline \eta = 1$. As seen from Fig. \ref{evdbn}, the cumulative 
distribution of $\eta$ for both the cases have very similar behaviour. 
Considerable deviations from the RMT predicted cumulative Porter-Thomas 
distribution, $I(\eta )= \hbox{Erf}\left(\sqrt{\eta/2}\right)$, may be 
attributed to the localization of the states. However, the distribution 
corresponds to $R=0.5$ tends to be closer to the RMT predicted behaviour.

\section{Summary and Conclusion}

A particle inside a one-dimensional infinite square well potential in the
presence of a time periodic pulsed field is examined both classically and 
quantum mechanically. This simple model can be seen as one generalization of 
the kicked rotor, or the standard map. A variety of classical dynamical 
features emerge from the nature of the ratio, $R$, of the two competing length 
scales (the well width and field wavelength). Many of the dynamical features 
so observed  are generic to a  wide class of systems of substantial current 
interest {\it viz.} externally forced particles in wells.

It is shown that when the length scales do not match, even in perturbative
regimes dynamics can be increasingly complex wherein all the KAM tori in phase 
space break up. As a result the transition to chaos is abrupt, a typical 
scenario of non-KAM behaviour. Quantum mechanically imprints of such a 
transition is seen as a spread in bulk part (delocalization) of the 
eigenstates. Thus we realize the length scale ratio $R$ as a control parameter 
for the localization in the weak field regime. 

On increasing the field strength, chaos assisted diffusion takes place in 
momentum. From earlier studies on the kicked rotor it is known that the 
average localization length of the eigenstates is directly proportional to 
the classical diffusion coefficient. We have shown that in our generalization 
of the kicked rotor also, this result grossly explains the localization 
behaviour of the eigenstates through the classical transport property. Thus 
the kicked rotor continues to serve as a useful model in understanding 
physical phenomena exhibited by a larger class of systems. 

We have observed, as expected, that in the regular case the nearest neighbour 
spacing distribution of the quasienergies show good agreement with Poisson 
distribution. We have presented evidence to support that, in highly chaotic 
regimes spacings show some deviations from the Poisson distribution though 
the corresponding eigenstates belong to an unbounded phase space. Limited 
overlap of the eigenstates results in such deviations. However, spacing does 
not show the RMT predicted Wigner distribution as was claimed in an earlier 
study \cite{bambi99}. The earlier result is attributed to lack of converged 
states which make up the statistics. 

While spacing is not much sensitive to the classical chaos, the distribution 
of participation ratios of the eigenstates is shown to be a good measure 
to distinguish chaotic quantum systems from the regular ones. Quantum 
mechanically, chaotic regimes are characterized by a lognormal distribution 
of the participation ratios. In addition to the above generic quantum features,
we have also studied non-generic phenomena like ``quantum resonance''. In 
resonance condition, kinetic energy of the particle grows quadratically with 
the number of kicks. This unbounded energy growth is not affected by the 
length scale ratio and can enhance ionization in the finite well system. 

As far as experimental realizations of this work is concerned, perhaps both 
quantum wells \cite{wilk96} and cold atom experiments \cite{moore95} are 
possible candidates. As suggested above, the $R$ effects may be best observed 
at small field strengths and for $R>1$. Further work is underway exploring the 
nature of localization in such systems, including a bounded version of the 
generalized standard map.

\newpage

\end{document}